%% file: itw.20160315.tex
\documentclass[preprint2]{aastex}

\shorttitle{PARALLAX RESULTS FROM URAT EPOCH DATA}
\shortauthors{Finch \& Zacharias}

\begin{document}

\title{PARALLAX RESULTS FROM URAT EPOCH DATA}
\author{Charlie T. Finch, Norbert Zacharias}

\email{finch@usno.navy.mil}

\affil{U.S. Naval Observatory, Washington DC 20392--5420}

\author{}
\affil{}


\begin{abstract}

We present 1103 trigonometric parallaxes and proper motions from the
United States Naval Observatory (USNO) Robotic Astrometric Telescope
(URAT) observations taken at the Naval Observatory Flagstaff Station
(NOFS) over a 3 year period from April 2012 to June 2015 covering the
entire sky north of about $-10^{\circ}$ declination.  We selected 2
samples: previously suspected nearby stars from known photometric
distances and stars showing a large, significant parallax signature in
URAT epoch data wihtout any prior selection criteria.  All systems
presented in this paper have an observed parallax $\ge$ 40 mas with no
previous published trigonometric parallax.  The formal errors on these
weighted parallax solutions are mostly between 4 and 10 mas.  This
sample gives a significant (order 50\%) increase to the number of
known systems having a trigonometric parallax to be within 25 pc of
the Sun (without applying Lutz-Kelker bias corrections).  A few of
these are found to be within 10 pc.  Many of these new nearby stars
display a total proper motion of less than 200 mas/yr.  URAT parallax
results have been verified against Hipparcos and Yale data for stars
in common.  The publication of all signifigant parallax observations
from URAT data is in preparation for CDS.


\end{abstract}


\section{INTRODUCTION}
\label{sec:introduction}

Due to the proximity to the Sun, nearby stars are excellent candidates
for multiplicity, stellar activity, ages and exoplanet research.  
A comprehensive census of nearby stars is needed to investigate the
luminosity and mass function of the solar neighborhood.

Trigonometric parallax is the most direct and accurate method to determine
stellar distances requiring only the Earth's orbital motion and 
multiple epoch astrometric observations of the target star.
This greatly reduces confusion or bias that may be present in other 
methods that rely on specific properties or prior knowledge about the star.

The primary source for trigonometric parallaxes are the Yale Parallax
Catalog (YPC) \citep{YPC} containing 8112 stars with a faint limit of
V = 21.5 magnitudes and the Hipparcos catalog \citep{HIP1} containing
118218 stars with a faint limit of V $\approx$ 13 mag.  The Hipparcos
new reduction \citep{HIP2} contains 117955 stellar parallaxes.  Due to
the magnitude limit of the Hipparcos catalog with completeness between
V 7.3 and 9.0 mag (dependent on galactic latitude and spectral type)
\citep{perryman} and the limited number of stellar parallaxes from
YPC, many nearby stars have yet to be discovered. Some recent ground
based efforts include the Research Consortium On Nearby Stars (RECONS)
\citep{TSNXIII,TSNXVII,TSNXXI,TSNXXII,TSNXXXIII,TSNXXIV}, USNO
\citep{dahn1,vrba1}, and MEarth \citep{mearth} along with many others
which have been trying to fill in the gaps before Gaia data becomes
available.

For this study we take the recently completed Northern Hemisphere
observing from the United States Naval Observatory (USNO) Robotic
Astrometric Telescope (URAT).  This data include all individual
exposures from April 2012 to June 2015 giving a longer epoch baseline
for determining parallaxes over the 2-year span of the First USNO
Robotic Astrometric Telescope Catalog (URAT1) \citep{URAT1} published
data (see $\S$~\ref{sec:OBS}).  We use previous proper motion surveys
utilizing the Digitized Sky Survey (DSS) in the northern sky
\citep{Lepine1}, the SuperCOSMOS Sky Survey in the southern sky
\citep{TSNXV,TSNXXIII,TSNXXV,TSNXXXV}, and the USNO CCD Astrograph
Catalog (UCAC) \citep{upm1,upm2,ucac4s} to identify stars previously
anticipated to be within 25 pc using other methods, like photometric
distances(see $\S$~\ref{sec:PRE}).  For these stars we obtain
trigonometric parallaxes for the first time using URAT data.

In addition, we also use the entire URAT Northern Hemisphere
observations to look for previously unknown nearby stars within 25 pc
of the Sun, independent of any selection criteria other than the
trigonometric parallax signature in the URAT data
(see $\S$~\ref{sec:NTPWPS}).  This task is difficult due to the relatively
short epoch span and moderate accuracy of our URAT observations and
the presence of contamination from e.g.~unknown multiplicity of stars
and statistical biases.

\section {OBSERVATIONS}
\label{sec:OBS}

All URAT sky survey observations were taken close to the meridian
(hour angle typically within $\pm 5^{\circ}$).
Each observing night is split into 5 equal long periods during
which a different set of 3 dither positions of a field are
observed, thus providing good parallactic angle distribution over
the course of a year.  
A 60 and 240 second exposure is taken at each individual telescope
pointing.  The entire pattern is repeated with 10 and 30 second
exposures with an objective grating near full moon.  This vastly
expands the dynamic range of the URAT survey toward bright stars,
covering the entire range between about R = 4 and 18 magnitudes. 
URAT observes through a single filter (part of the dewar window)
to provide a fixed bandpass of about 680 to 760 nm.

The clear aperture of the USNO astrograph is 206 mm with a focal
length of only 2 m.  A single exposure covers 28 square degrees with a
resolution of 0.9 arcsecond/pixel.  Each of the 4 large CCDs in the
focal plane covers a 2.65 by 2.65 deg area on the sky.  Data of all 3
years of operations (April 2012 to June 2015) at the USNO Flagstaff
Station (NOFS) are used here for this parallax investigation.  For
more details about the project, instrument and observing the reader is
referred to the URAT1 paper \citep{URAT1}.

\section {ASTROMETRIC REDUCTIONS}
\label{sec:RED}

\subsection {Raw Data Processing}

Custom code was used to apply overscan, trim, dark and flat-field
corrections operating on 2-byte-integer FITS files for both the
raw data and processed images. 
Custom code was also used to detect stellar images (4-sigma above background
threshold) and perform 2-dim spherical symmetric Gaussian model profile
fits to the observed stellar images of the processed pixel data.
Besides the seeing, a significant contribution to the observed point 
spread function (PSF) comes from diffraction due to the relatively small 
aperture, leading to an observed image profile width of typically 2 pixels
full width at half maximum (FWHM).  The PSF is very uniform across the
entire field of view, thus the same model function was used independent
of the location of a stellar images in the focal plane.

\subsection {Astrometric Solution}

An 8-parameter 'plate' model was adopted for the astrometric
reductions (linear + tilt terms) using the Fourth USNO CCD Astrograph 
Catalog (UCAC4) \citep{UCAC4} for reference stars, restricted to 
magnitudes R $=$ 8 to 16.  A weighted, least-squares solution with
outlier rejection was performed on each individual CCD and exposure
with typically several hundred to a few thousand reference stars per
astrometric solution.
The data are corrected for geometric field distortions (about 10 to 60 mas
effect) and pixel phase errors (0 to 15 mas) using 'look-up' tables
generated from preliminary reductions and residual analysis.  

Individual epoch
positions ($\alpha, \delta$) of all stars were obtained on the
International Celestial Reference System (ICRS) via UCAC4. Depending on
brightness and exposure time typical positional errors of individual
observations are 10 to 60 mas. All individual epoch positions have
been matched to individual stars with mean data and indexing saved
to a separate, large file allowing fast, direct access to the individual
observations (42 bytes of binary data each).
These URAT data consists of over 392 million individual objects with
about 8.4 billion observations.  Excluding single detections, the
average number of observations per star is about 24.
For more details about the astrometric reductions the reader is
referred to the URAT1 paper \citep{URAT1}.

\subsection {Solving For Parallax}

The URAT parallax pipeline utilizes routines from \citep{jaothesis},
the JPL DE405 ephemeris and Green's parallax factor \citep{green} for
determining parallaxes.  We first determined the location of the Earth
(rectangular coordinates $X$, $Y$, and $Z$) at each URAT mid exposure
using the JPL DE405 ephemeris.  This was done so that the pipeline can
be run over all epoch data of all stars and look up the Earth's
position by exposure number instead of re-calculating these data for
each star and epoch, thus significantly reducing the run time of the
code. The rectangular coordinates at epoch are then used to determine
the parallax factors using the following formulae:

\begin{equation}
P_{\alpha} \ =  \ X \sin \alpha \ - \ Y \cos \alpha 
\end{equation}
\begin{equation}
P_{\delta} \ = \ X \cos \alpha \sin \delta \ + \ Y \sin \alpha \sin \delta
 \ - \ Z \cos \delta 
\end{equation}

The parallax factors of all individual data are then used to
simultaneously solve for mean position, proper motion and parallax using
all 'good' epoch data of a given star in a least-squares adjustment
with outlier rejection from the following equations:

\begin{equation}
x(t) \ = \ x(t_{0}) \ + \ \mu_{x} (t - t_{0}) \ + \ \pi P_{\alpha}
\end{equation}
\begin{equation}
y(t) \ = \ y(t_{0}) \ + \ \mu_{y} (t - t_{0}) \ + \ \pi P_{\delta}
\end{equation}

Where $x(t), y(t)$ are the positions of a given star on the tangential
plane as function of time ($t$), $\pi$ is the parallax, $\mu_{x} =
\mu_{\alpha} cos\delta$ and $\mu_{y} = \mu_{\delta}$ represent the
proper motions in RA and DEC, respectively.  The initial instant of
time ($t_{0}$) can be chosen arbitrarily.  Here we choose the first
observing epoch as zero point for $x, y$ and $t$.

In Table~\ref{table_cuts} we show the cuts we have adopted to the URAT
epoch data of a given star when solving for parallax. The limits imposed
on individual image amplitude, image profile width (FWHM) and position
fit error (sigma) are set to not allow saturated stars, stars with
too few photons or poorly determined positions to be used in the
parallax solution.  Limits on number of individual observations used 
and epoch span were adopted empirically to obtain 'good' results when
comparing the URAT parallaxes with the Hipparcos Catalog and the 
Yale Parallax Catalog and at the same time allow for as many as
possible 'reasonable' parallax solutions from our data.

In the least-squares adjustment to solve for parallax (eqs.~3 and 4)
weights were used according to the precision of individual
observations which can vary largely due to the different exposure
times and amplitudes of individual observations.  An error floor of 10
mas was added in quadrature to the random errors (in the astrometric
reductions of individual observations) before calculating the weights
for an observation in determining the parallax (see \citep{URAT1} for
details about the astrometric reductions of individual observations).
A 3-sigma outlier rejection criteria was imposed and in cases where
the post-fit error of unit weight exceeded 1.5 times the expected
error of the fit solution the largest residual(s) were iteratively
removed from the fit solution.  This way typically up to a few percent
of observations were rejected for the parallax solution of individual
stars.

\subsection {Conversion From Relative To Absolute Parallax}

At this point we have parallaxes for the target stars that are
relative to the set of reference stars used in the reductions of the
URAT positions, i.e.~with respect to typically many hundreds to a few
thousand stars in a 2.65 by 2.65 square degree area of a single URAT
CCD detector.  To convert from relative to absolute parallax we use
photometric parallaxes for the reference stars.  This method is not as
reliable as spectroscopic parallaxes which can determine the spectral
type and luminosity class, allowing to apply correct Mv-color
relations to each reference star.  
Photometric parallaxes were used because of the lack of spectroscopic
data for millions of stars in our survey area down to R = 16 mag.

Again we use the UCAC4 catalog here now also for its photometric data.
This was done by picking the same UCAC4 stars previously used as
reference stars in the astrometric reductions (R = 8 to 16 mag).
These stars are then run through our 16 photometric color-$M_{K{_s}}$
relations \citep{ucac4s} which use the AAVSO Photometric All-Sky
Survey (APASS) $BVgri$ and Two Micron All Sky Survey (2MASS) $JHK_s$
photometry included in the UCAC4 catalog. Using this data we determine
a mean photometric parallax per reference star, assuming all stars are
main-sequence due to the lack of information for each star.  This is
done for all reference stars in a 2 by 2 degree wide area around each
target star to determine a mean absolute parallax correction for each
target star.  The average parallax correction for the target stars in
this paper is 1.3 mas varying from 0.5 to 6.6 mas depending on the
field.  For any star having an unknown correction, we use the mean to
convert to absolute parallax.  For stars having a correction greater
than three times the average ($\ge$ 3.9 mas) we adopt this as a cutoff
and use 3.9 mas to convert to absolute parallax.  For these stars we
have added a note in the tables to indicate if the mean or cutoff
correction was used.  Considering the relatively large random errors
of our parallax results, this step is not critical for the goal of our
investigation.

\subsection{Biases}

We do not apply any corrections to our parallaxes for the Lutz-Kelker
bias \citep{LK73} because we don't draw conclusions about completeness of 
a distance limited sample or interpret results regarding absolute luminosity.
This is beyond the scope of this paper which is to present the observed
trigonometric parallaxes for a large number of stars.
This bias comes into play when "translating" the observed parallax 
into distances and absolute luminosities.


\section {RESULTS}

The URAT parallax pipeline as described above was run over the entire
set of URAT observational epoch data resulting in 44.7 million
parallax solutions.  These results were then matched with the
Hiparrcos catalog, YPC, SIMBAD database \citep{simbad} and other
trigonometric parallax published data using a 60 arcsecond radius 
to flag stars with previously published trigonometric parallaxes. 
Results for nearby stars and no prior published trigonometric parallax 
were extracted from this pool of data, while for most of those millions
of stars our parallax solution is not significant due to the large
distance of most stars and our relatively large random observational
errors.

\subsection {Comparison To Published Parallaxes}

The Hipparcos Catalog contains 65546 stars north of $-10^{\circ}$
declination.  While the URAT catalog goes as far south as about
$\delta = -24^{\circ}$ its completeness begins to drop south of
$-10^{\circ}$ declination.  The URAT catalog recovers 65524 (99.96\%)
of Hipparcos stars in this area of the sky.  Most of the few missed
stars are just brighter than the URAT limit or lost due to blended
images which are not taken care of in the URAT reductions.  

One of the main goals of this paper is to discover new nearby stars or
confirm candidates within 25 pc.  To determine how well URAT can detect 
stars within 25 pc we used the Hipparcos catalog (2007 version).  Hipparcos
contains 926 stars north of $-10^{\circ}$ declination with a parallax
$\ge$ 40 mas.  URAT recovers 887 (= 95.8\%) of the Hipparcos 25 pc
sample loosing 20 from our amplitude/sigma cuts, 16 having too short
of an observational epoch span, 3 having too few observations and 6 stars
not being in the URAT data at all.  Of those 887 recovered stars, 778
(= 87.7\%) have a URAT parallax $\ge$ 32 mas. 
In Figure~\ref{hip25urat} we show the comparison between the Hipparcos 
25 pc sample and URAT results, where the center line indicates agreement
and the 2 outer lines indicate a 10 mas difference from the Hipparcos
parallax values.

In order to investigate how reliabel the URAT parallax formal errors
are, we took the 25 pc sample of 887 stars common to Hipparcos and
applied 2 different cases of cuts to filter "reasonably good" data.
Selection criteria and results are summarized in Table~\ref{table_err}. 
The number of
observations per target star, the epoch span of these observations and
the formal errors of both the Hipparcos and our URAT data were
restricted.  For the stars remaining in the sample the unweighted
difference in parallax was calculated.  A few outliers were further
eleminated at this stage (see Table~\ref{table_err}) before the RMS of
the parallax difference and the formal errors were calculated.  We see
that formal errors on our URAT parallaxes are underestimated by about
10 to 20\% when compared to the observed scatter of the data (the
errors for the Hipparcos parallaxes are much smaller, typically 0.5 to
2 mas, than the URAT parallax errors). Distributions of the URAT
parallax formal errors are shown in Figures~\ref {hist_hip_st} and
\ref{hist_hip} for the Hipparcos 25 pc sample and all Hipparcos stars
north of $-10^{\circ}$ declination, respectively. This shows that the
URAT errors for these samples peak around 8 mas with typical errors 
between 4 and 15 mas.

In Figure~\ref{uraterr}, we show the relation between the URAT 
parallax error and epoch difference of the available URAT data for 
the Hipparcos 25 pc sample.  For Figure~\ref{uratnp}, we use the same
sample but show the relationship between the URAT parallax error 
and number of observations.  These plots indicate that 
the URAT parallax errors drop below about 8 mas for solutions 
having an epoch span greater than 2.0 years and more
than 40 individual epochs.

Hipparcos and YPC have 698 stars in common north of $\delta =
-10^{\circ}$ with a V magnitude range of 2.83 to 11.70.  Of these 698
stars, we obtained a URAT parallax for 696 stars.
Figures~\ref{urat.hip}, \ref{urat.ypc} and \ref{hip.ypc} show the
comparison of URAT vs.~Hipparcos, URAT vs.~YPC and YPC vs.~Hipparcos,
respectively.  The center line indicates perfect agreement while the two
outer lines have been added to show a difference of 10 mas.  URAT
compares relatively well with both the Hipparcos and YPC results, but
does show a somewhat larger scatter not present in the YPC vs.~Hipparcos
plot, as expected from mean formal errors.

A similiar parallax investigation was completed using images from the
MEarth photometric survey \citep{mearth} having a large overlap with
our URAT parallaxes.  The typical error on the MEarth parallax data is
reported to be 4 mas, while that for URAT data is about twice as large. 
In Figure~\ref{urat.mearth}, we compare the parallax results reported in
\citep{mearth} for the 572 stars matched with the URAT parallax data.
The center line represents perfect agreement while the outer
lines show a $\pm$ 10 mas difference.  

In Figure~\ref{all_known} we show a comparison between URAT and 1387
published trigonometric parallaxes $\ge$ 40 mas from the 11878
parallaxes described in $\S$~\ref{sec:NTPWPS} pulled from the 44.7
million parallaxes from the entire URAT data.  A 60 arcsecond radius
was used when searching SIMBAD, published papers and vizeir for
published trigonmetric parallaxes.

\subsection {New Trigonometric Parallaxes of Photometric Sample}
\label{sec:PRE}

We first took a list of 9737 nearby stars having previously published
distance estimates based on photometry from the publications listed in
$\S$~\ref{sec:introduction}.  After cutting these to the area of sky
being searched in this survey ($\delta \ge -10^{\circ}$) we are left
with 4853 candidate nearby stars.  These are then run through our
parallax pipeline and list of constraints resulting in 3093 parallax
results.  These constraints include the same as listed in
Table~\ref{table_cuts} with an additional constraint to the parallax
error to be $\le$ one half the parallax and $\le$ 25 mas.  Of the
remaining 3093 stars, 1906 were found to not have a published parallax
result after searching SIMBAD, Vizier and published papers.  After
removing 182 duplicate entries which came from combining all systems
from the published papers we are left with 1724 parallax results.  We
then only keep the 545 stars having a URAT parallax $\ge$ 40 mas which
is the focus of this paper.  Of these 545 stars 10 have a URAT
parallax $\ge$ 100 mas.

Data for all these 545 stars (picked from photometric list and having
a URAT prallax distance within 25 pc) are given in
Table~\ref{table_predist} (sorted by parallax).  We include the names,
RA and Dec coordinates (at epoch 2014.0 on the ICRS; derived from
input data in corresponding published paper), URAT magnitude, epoch
span, number of observations, number of rejected observations,
absolute parallax, parallax error, parallax correction, proper motion
in RA and Dec with associated errors, and $JHK_sBVgri$ photometric
data from UCAC4 along with any notes.  All 545 stars have no previous
published trigonometric parallax.  The average formal parallax error
for this sample is 7.8 mas. Notes on specific individual stars from
this sample are reported in $\S$~\ref{sec:NOIS}.


In Figure~\ref{photcom}, we compare the URAT trigonometric parallax to
the previously published photometric parallaxes for the 545 stars in
this sample.  In Figure~\ref{goodfit}, we show an example of a good fit
for G165-058 ($\pi$ = 80.1 $\pm$ 4.4 mas, pmra = 6.4 $\pm$ 2.8 mas/yr, 
pmdc = 298.1 $\pm$ 2.7 mas/yr).  This plot shows all URAT epoch data for
this star. 
A histogram of the parallax error for this sample is given in
Figure~\ref{pi-pre-perr}, showing the peak around 6 mas with the
majority of the errors falling in the 4 to 10 mas range.  In
Figure~\ref{pi-pre-tpm} we present a histogram of total proper motions
for the same sample which shows a significant number of nearby stars
with small proper motions (between 100 and 450 mas/yr).  The errors
for this sample are somewhat smaller than for the Hipparcos and YPC
sample used above due to the more stringent restrictions applied here.

\subsection {New Trigonometric Parallaxes Without Prior Selection}
\label{sec:NTPWPS}

The entire URAT data was run through the parallax pipeline giving 44.7
million parallax solutions.  We then implemented several cuts to get
this to a more manageable sample, keeping only stars with:
1) a parallax $\ge$ 40 mas, 2) epoch span
of at least 1.5 years, 3) at least 30 individual epochs, 4) parallax
error $\le$ 20 mas and $\le$ one-fourth the absolute parallax.  These cuts
left us with a sample of 6571 new candidate nearby stars.  We then
used SIMBAD, Vizeir and published papers to remove 1753 stars having a
previous published parallax or having a match with
Table~\ref{table_predist}, leaving us with 4818 nearby star
candidates.  Of these new nearby star candidates 35 have a URAT
parallax $\ge$ 100 mas.

The current URAT reduction process does not take provisions for close
doubles (blended images) of arcsecond-level separations. 
Many of our candidates, particularly those with large parallax solution
errors are possibly blended images.  This means a visual inspection
(residual plots as well as real sky image using Aladin) must be 
completed to verify the integrity of the solution.  This would not 
be practical for the entire sample.

However, this visual inspection was completed for the 35 candidate
stars in our 10 pc sample.  For this sample we find the majority of
the stars are blended or elongated even if the fit solution looks
reasonable.  All stars considered 'suspect' from this visual
inspection are flagged.  In Figure~\ref{suspectfit}, we show an example
of a poor fit case for the star UPM 1304+8611 
($\pi$ = 42.5 $\pm$ 7.3 mas, pmra = 40.1 $\pm$ 8.6 mas/yr, 
pmdc = -33.4 $\pm$ 8.4 mas/yr), which is marked
as suspect in our table.  This plot shows all URAT epoch data for this
star similarly to the previous Figure.

While sifting through the 10 pc nearby star sample we notice
UPM2340+4631, and UPM2340+4625 both have large parallaxes with a
separation less than 7 arcminutes.  We also notice that
UPM1758-0055, UPM1758-0057, UPM1758-0058 and UPM1758-0104 also share a
large parallax and are all separated by less than 12 arcminutes,
indicating they may all be part of the same group.  
More information on these stars are reported in $\S$~\ref{sec:NOIS}.

For the remaining 4783 stars comprising of nearby star candidates
between 10 and 25 pc, we were not able to search the entire sample.
Instead we sorted the list by 1) number of epochs, 2) parallax error
and 3) image elongation in that order.  We then started visually checking
the fits to verify the quality.  During this process we also
randomly looked at sky images using Aladin.  For this sample we did
not see any examples of an elongated or blended image in our random
search.  The fits for this sample also seemed of higher quality than
the 10 pc sample however, we did see some fits which might be suspect
which are noted in the table. We were able to search 1221 of the 25 pc
nearby star candidates that we felt had the highest chance of being a
real nearby star with a quality parallax solution.

This search produced 558 new discoveries within 25 pc of the Sun, of
which 24 have a URAT parallax putting them within 10 pc.  All of the
fits and sky images have been visually inspected for the 10 pc sample.
For new discoveries between 10 and 25 pc all fits have been
visually inspected and sky images randomly inspected for issues.


In Table~\ref{table_newdist}, we present details for the 558 new nearby
stars having an absolute parallax $\ge$ 40 mas (sorted by parallax).
These stars have no previously published photometric distance or
trigonometric parallax information.  We have added a note in the table
to indicate if the sky image was found to be elongated, blended or the
fit suspect.  We include the names, RA and Dec coordinates (ICRS epoch
2014; derived from the mean URAT position at mean epoch from all
epochs used in the fit along with the URAT proper motions), URAT
magnitude, epoch span, number of observations, number of rejected
observations, absolute parallax, parallax error, parallax correction,
proper motion with associated errors, and $JHK_sBVgri$ photometric
data from the URAT1, along with notes.  We have given a USNO Proper
Motion (UPM) name for the 281 entries having no previous proper motion
information reported from SIMBAD. The average absolute parallax error
for this sample is 6.3 mas. Details about individual stars are
reported in $\S$~\ref{sec:NOIS}.

We present a histogram of the parallax error for this sample in
Figure~\ref{pi-ndw-err}, showing the peak around 5 mas with the
majority of the errors falling in the 4 to 8 mas range.  In
Figure~\ref{pi-ndw-tpm} we show a histogram of total proper motions
for the same sample which shows the majority of the new nearby star
discoveries have small proper motions (less than 200 mas/yr.  The
errors for this sample are the smallest out of any other sample due to
the most stringent selection criteria applied here for searching the 
best possible nearby star candidates.

\section {NOTES ON INDIVIDUAL SYSTEMS}
\label{sec:NOIS}

{\bf TYC 3980-1081-1} having a V magnitude of 10.5 is the closest
confirmed star in this sample at 6.46 pc ($\pi$ = 154.8 $\pm$ 12.1
mas, pmra = -21.0 $\pm$ 13.6 mas/yr, pmdc = 149.5 $\pm$ 13.7 mas/yr),
making this the 97th nearest star system to the Sun. The new URAT
parallax is slightly farther than the photometrically estimated
distance of 5.9 pc from the UCAC4 nearby star survey paper
\citep{ucac4s}.  This star shows no signs of elongated or blended
image and no issues are seen with the fit solution.  Our measurements
suggest that this would be the 97th closest star system according to
the Research Consortium On Nearby Stars (RECONS)
\footnote{\it www.recons.org} 100 nearest star systems.

{\bf UPM 0012+5215} having a URAT magnitude of 15.62 is the third
closest new discovery in this sample at 6.48 pc ($\pi$ = 154.3 $\pm$
17.3 mas, pmra = 42.6 $\pm$ 14.5 mas/yr, pmdc = 46.2 $\pm$ 14.0
mas/yr). Our measurements suggest that this would be the 99th nearest
star system to the Sun.  The sky image of this star did show a blended
image and has been noted as suspect in Table~\ref{table_newdist},
indicating the solution may be erroneous.

{\bf UPM 2340+4625} having a URAT magnitude of 14.93 is the second
closest new discovery in this sample at 4.8 pc ($\pi$ = 206.5 $\pm$
17.3 mas,pmra = 48.5 $\pm$ 12.3 mas/yr, pmdc = 42.0 $\pm$ 12.0
mas/yr).  Our measurements suggest that this would be the 44th nearest
star system to the Sun.  The sky image of this star does show an
elongated image and the solution has been noted as suspect in
Table~\ref{table_newdist}.  However this star shares a large parallax
with {\bf UPM 2340+4631} (within 2 sigma of each other) at a
separation less than 7 arcminutes away indicating it may be a binary.
The sky image of UPM 2340+4631 does show an elongated image and has
been noted as suspect in Table~\ref{table_newdist}, indicating the
solution may be erroneous.


{\bf UPM 0429+3806} having a URAT magnitude of 15.36 is the closest
new discovery in this sample at 4.4 pc ($\pi$ = 227.2 $\pm$ 19.0 mas,
pmra = 114.7 $\pm$ 14.0 mas/yr, pmdc = 3.8 $\pm$ 13.0 mas/yr).  Our
measurements suggest that this would be the 34th nearest star system
to the Sun. The sky image of this star did show an elongated image and
has been noted as suspect in Table~\ref{table_newdist} indicating the
solution may be erroneous.

{\bf UPM 1758-0058, UPM 1758-0057, UPM 1758-0055 and UPM 1758-0055}
all share a large parallax ($\pi$ =
 103.2 $\pm$ 18.4 mas, 
 110.6 $\pm$ 18.9 mas,
 103.5 $\pm$ 22.0 mas,
 109.5 $\pm$ 23.3 mas)
within 1 sigma respectively at a separation less than 12
arcminutes on the sky and all with small proper motions.  
None of the sky images for these stars have been
tagged as having a blended or elongated image.  However these stars
have been tagged as suspect in Table~\ref{table_newdist} due to scatter
seen in the fits indicating the solutions may be erroneous.

\section {CONCLUSIONS}

Parallaxes from URAT data are as good as can be expected considering
that this is a wide-angle, all-sky survey using a telescope with only
2 m focal length and relatively short epoch span of about 3 years.
This is the largest trigonometric parallax survey conducted since the
Hipparcos mission and before PanSTARRS and Gaia results become available.

Contrary to dedicated trigonometric parallax programs the observations
of all target stars of our URAT data are scattered all over the focal
plane on different CCDs and different areas of the CCDs, even with
largely different image amplitudes due to the use of 10 to 240 sec
exposures.  Thus potential magnitude dependent systematic errors and
geometric field distortions must have been successfully removed from
the URAT data to insignificant levels, else the obtained parallax
results would have shown larger errors.

Although a formal trigonometric parallax solution from URAT data could
be obtained for over 40 million stars, only a small fraction of those
are significant due to the modest precision, limited number of
observations and epoch span and lack of double star fitting in the
URAT data.  However, many accessible stars within 10 and 25 pc can
reliably be detected independent of any pre-selections, just based on
the observed trigonometric parallax signature in the URAT data.

A large number (545) nearby star candidates in the sky area north of
$\delta = -10^{\circ}$ have been confirmed with URAT trigonometric
parallaxes to be within 25 pc of the Sun.  A significant number (558)
of new nearby stars have been discovered from this survey.

In Figure~\ref{sky}, we show the location on the sky of the entire 
1103 star sample from Tables~\ref{table_predist} and \ref{table_newdist}.
The distribution is relatively uniform over the area of our survey
showing no 'clumps'.

There is a significant overlap of targets between the MEarth project
and the URAT survey, however, the MEarth project is limited to very
red, pre-selected M dwarfs, while the URAT survey has no such selection
bias.  This explains the fair number of additional nearby stars discovered
in the URAT data.

Interestingly in Figure~\ref{pi-ndw-tpm}, we show a significant
number of nearby stars with small proper motions (less than 200 mas/yr
often adopted for high proper motion stars) indicating more nearby
stars hiding in the slower proper motion regime.

As of 01 January 2014, there are 2168 systems known all-sky within 25 pc,
of which 270 systems are known to be within 10 pc that have
accurate trigonometric parallaxes \citep{ucac4s}. The total number
(1103) of confirmed and new nearby stars from our URAT survey constitutes
a total increase of 51\% of the previously known nearby star sample
in this distance range and sky area. The 33 stars with a reported URAT
parallax $\ge$ 100 mas in this paper constitutes an increase of 12\%
of the previously known 10 pc sample.  All stars reported in this
survey have no previously known trigonometric parallax.
The real fraction of stars added to the 10 pc and 25 pc samples by our
survey will be somewhat smaller due to Lutz-Kelker bias and parallax
solution fit statistical bias.

A catalog of all significant parallaxes as obtained from URAT
regardless of prior publications is in preparation for CDS.
This will include our measures of stars also found in Hipparcos,
YPC, MEarth and other projects and will go beyond the 40 mas limit
adopted for this paper.



\acknowledgments

We thank the entire URAT team for making this nearby star search
possible.  Special thanks go to Wei-Chun Jao and members of the RECONS
team for help with the parallax pipeline.  This work has made use of
the SIMBAD, VizieR, and Aladin databases operated at the CDS in
Strasbourg, France.  We have also made use of data from 2MASS, APASS,
and the ADS service as well as the PGPLOT plotting software.  We also
would like to thank the many people who contributed to our custom
astrometric software code, some of which dates way back
e.g. {\citep{ausgl}.  We also benefit from the fact that Fortran code
  is backwards compatible allowing us to mix recent code with
  original, unchanged code written over the past 50 years.





  \begin{figure}
  \epsscale{1.00}
  \includegraphics[angle=-90,scale=1.0]{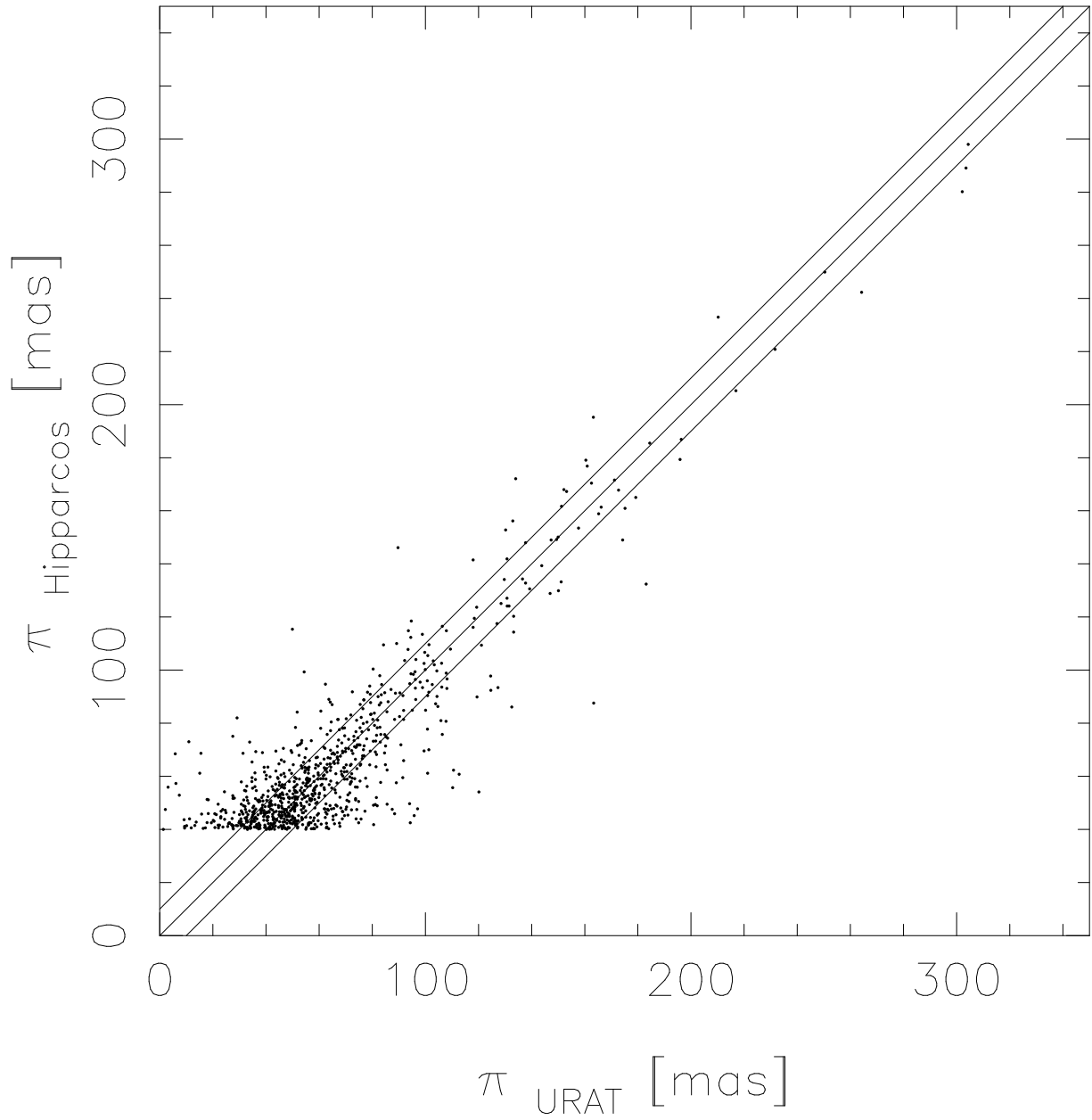}
  \caption{Parallax comparison between URAT and the Hipparcos 25 pc
    sample north of $\delta = -10^{\circ}$.  The center line
    represents perfect agreement while the outer lines show a $\pm$ 10
    mas difference.}\label{hip25urat}
  \end{figure}
  
 \clearpage
 
  \begin{figure}
  \epsscale{1.00}
  \includegraphics[angle=-90,scale=0.5]{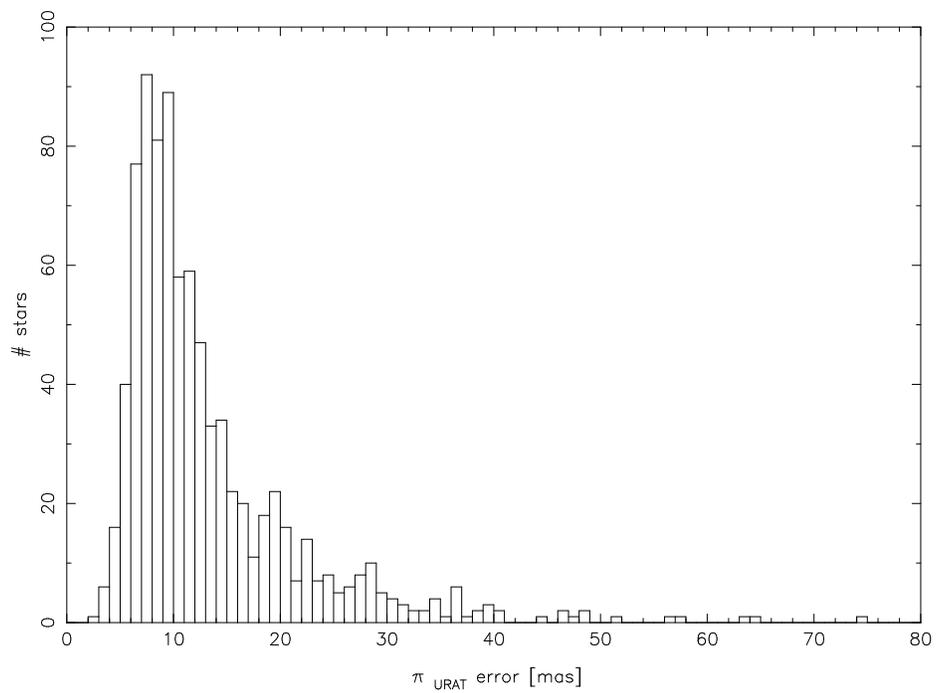}
  \caption{Histogram plot of URAT parallax errors for the
    stars in the Hipparcos 25 pc sample north of $\delta =
    -10^{\circ}$. }\label{hist_hip_st}
  \end{figure}
 
 \clearpage

  \begin{figure}
  \epsscale{1.00}
  \includegraphics[angle=-90,scale=0.5]{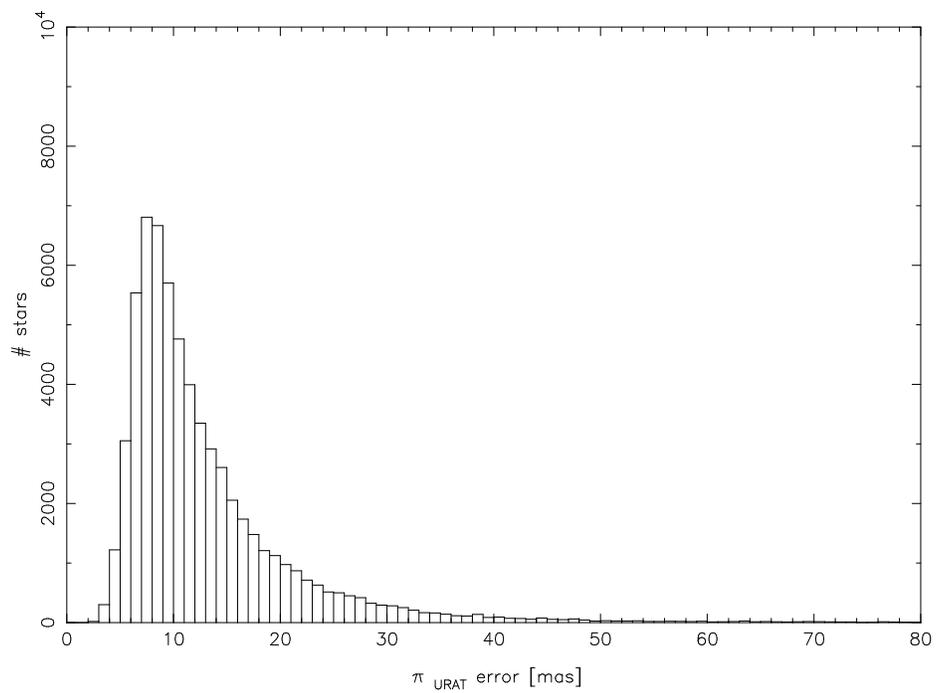}
  \caption{Histogram plot of URAT parallax errors for the
    63238 stars in the Hipparcos sample north of $\delta =
    -10^{\circ}$ }\label{hist_hip}
  \end{figure}
 
 \clearpage

  \begin{figure}
  \epsscale{1.00}
  \includegraphics[angle=-90,scale=0.5]{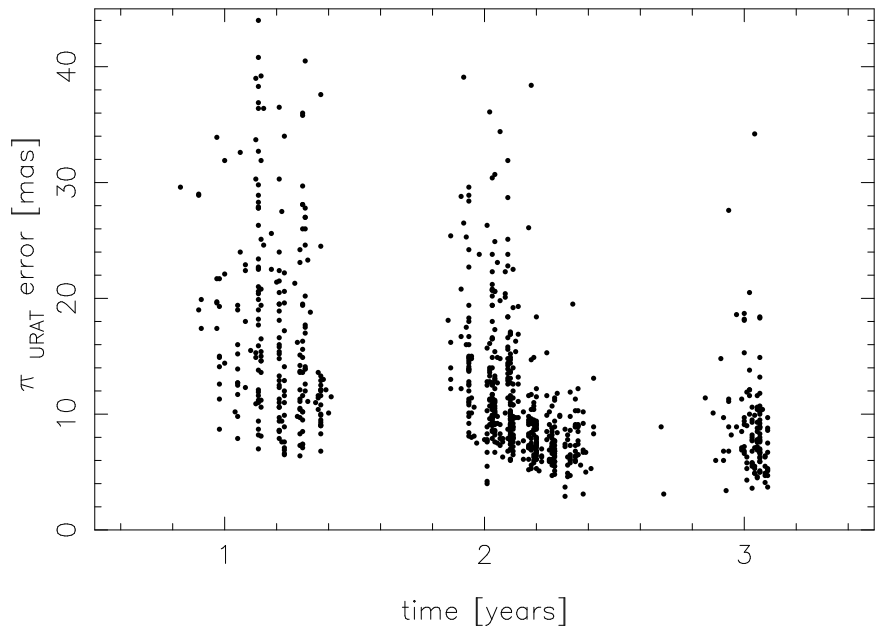}
  \caption{Relationship between URAT parallax errors and epoch span
    coverage for the Hipparcos 25 pc sample north of $\delta =
    -10^{\circ}$. }\label{uraterr}
  \end{figure}
 
 \clearpage

  \begin{figure}
  \epsscale{1.00}
  \includegraphics[angle=-90,scale=0.5]{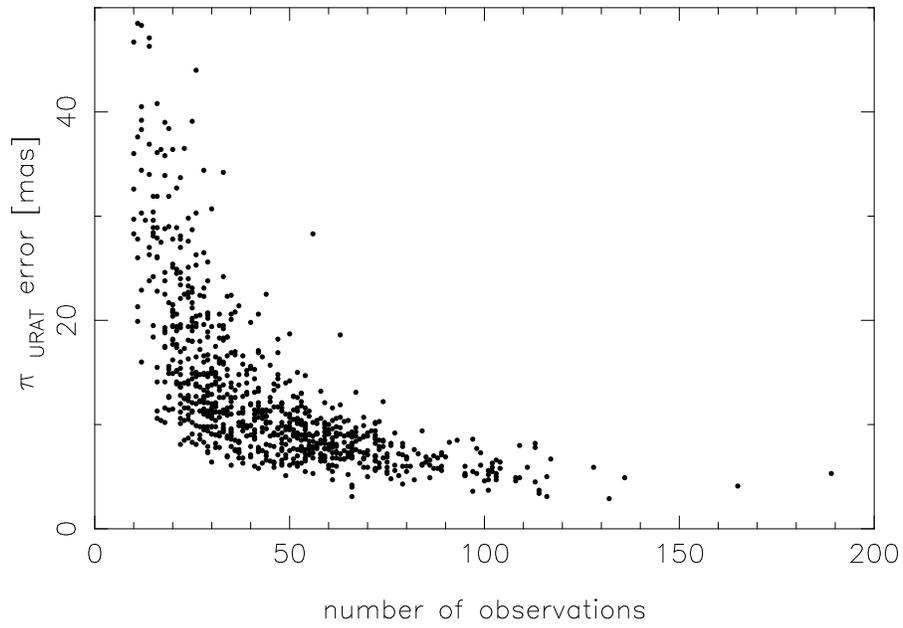}
  \caption{Relationship between URAT parallax errors and number of
    observations for the Hipparcos 25 pc sample north of $\delta =
    -10^{\circ}$. }\label{uratnp}
  \end{figure}
 
 \clearpage

  \begin{figure}
  \epsscale{1.00}
  \includegraphics[angle=-90,scale=0.5]{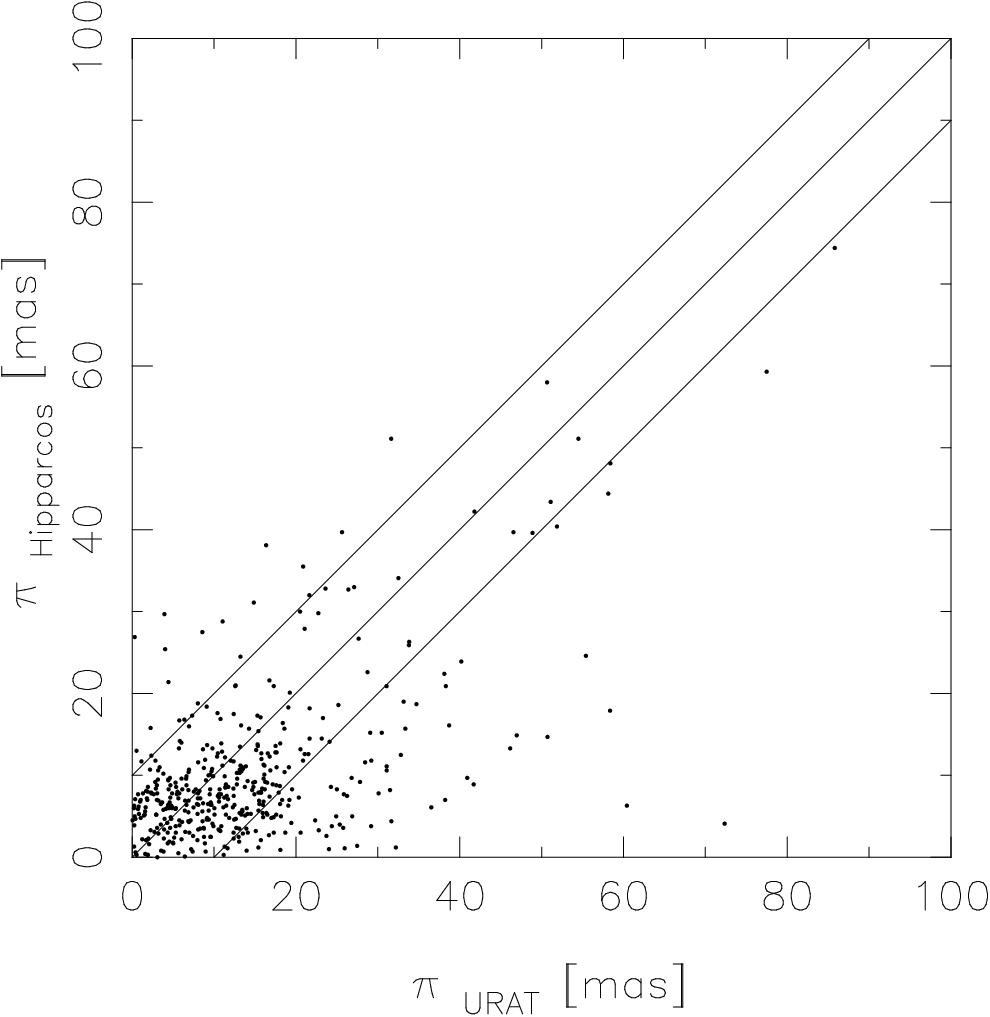}
  \caption{Comparison between URAT and Hipparcos parallaxes for stars
    in common with YPC north of $\delta = -10^{\circ}$.  The center
    line represents perfect agreement while the outer lines show a $\pm$
    10 mas difference. }\label{urat.hip}
  \end{figure}
 
 \clearpage

  \begin{figure}
  \epsscale{1.00}
  \includegraphics[angle=-90,scale=0.5]{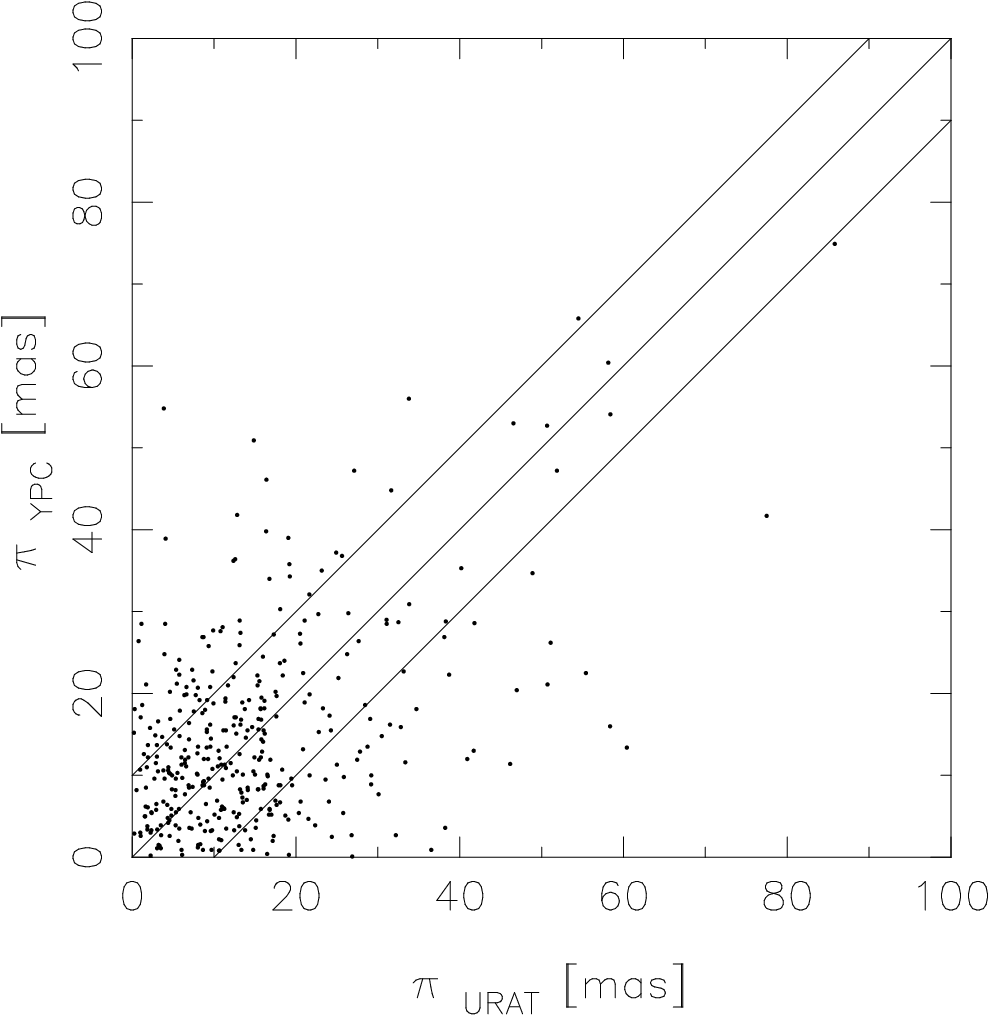}
  \caption{Comparison between URAT and YPC parallaxes for stars in
    common with Hipparcos north of $\delta = -10^{\circ}$.  The center
    line represents perfect agreement while the outer lines show a
    $\pm$ 10 mas difference. }\label{urat.ypc}
  \end{figure}
 
 \clearpage

  \begin{figure}
  \epsscale{1.00}
  \includegraphics[angle=-90,scale=0.5]{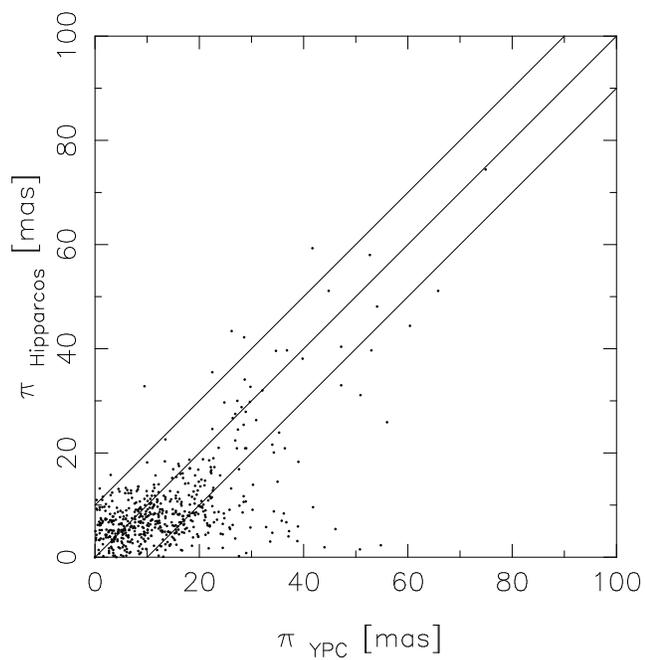}
  \caption{Comparison between YPC and Hipparcos parallaxes for stars
    in common with URAT north of $\delta = -10^{\circ}$.  The center
    line represents perfect agreement while the outer lines show a
    $\pm$ 10 mas difference.}\label{hip.ypc}
  \end{figure}
 
 \clearpage

  \begin{figure}
  \epsscale{1.00}
  \includegraphics[angle=-90,scale=0.5]{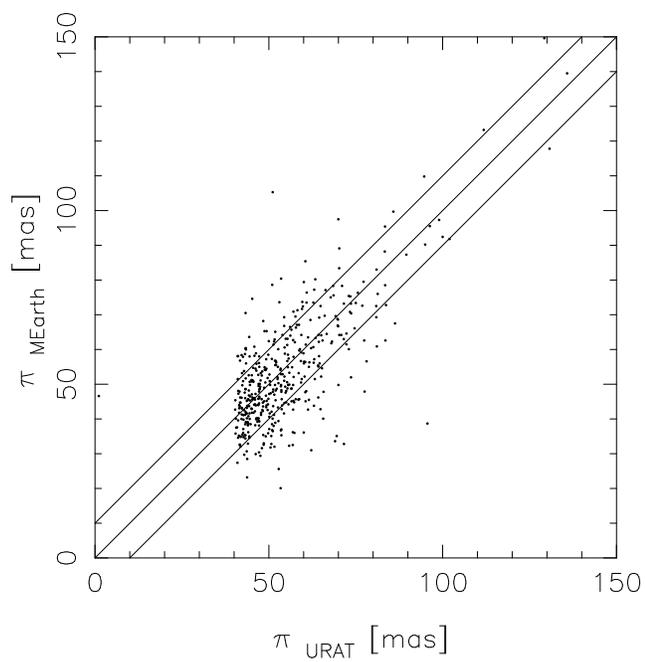}
  \caption{Comparison between URAT and the MEarth project \citep{mearth}
    parallaxes for the 572 stars in common.
    The center line represents perfect agreement while the outer lines 
    show a $\pm$ 10 mas difference. }\label{urat.mearth}
  \end{figure}
 
 \clearpage

  \begin{figure}
  \epsscale{1.00}
  \includegraphics[angle=-90,scale=0.5]{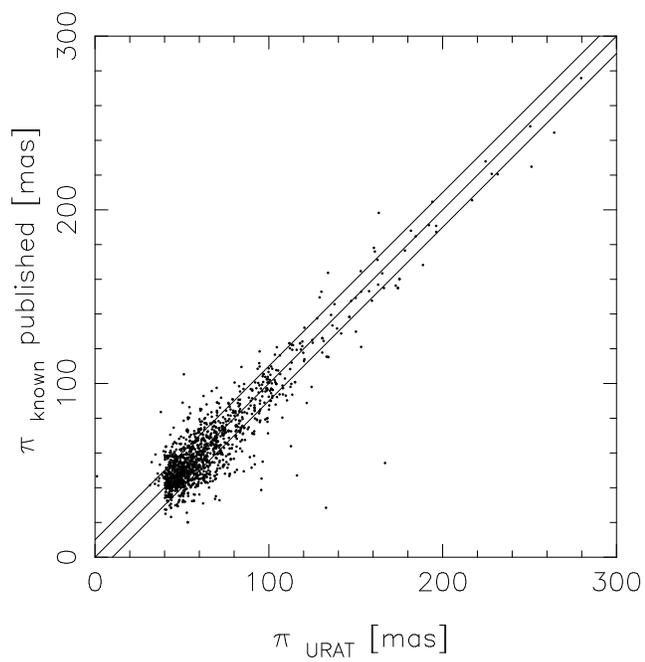}
  \caption{Comparison between the URAT 25 pc sample and published 
    trigonometric parallaxes of 1387 stars in common.  
    The center line represents perfect agreement while the outer 
    lines show a $\pm$ 10 mas difference. }\label{all_known}
  \end{figure}
 
 \clearpage

  \begin{figure}
  \epsscale{1.00}
  \includegraphics[angle=-90,scale=0.5]{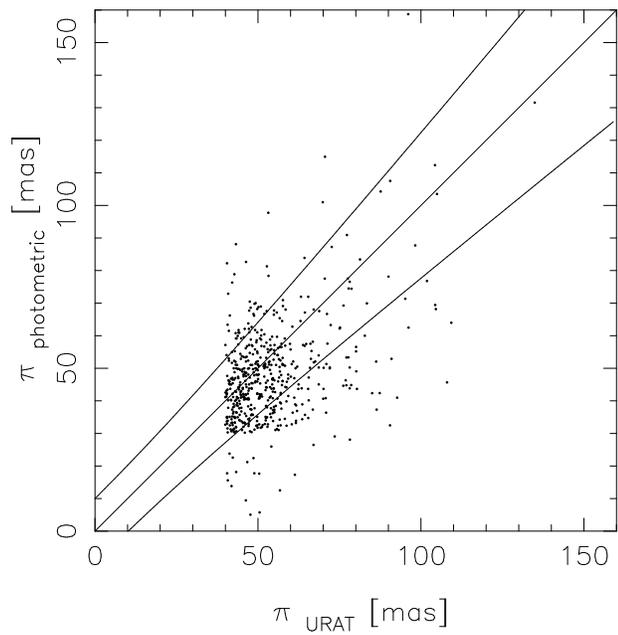}
  \caption{Comparison between the photometric and URAT parallax for
    the 545 stars without previously published trigonometric distance
    with $\pi \ge$ 40 mas and $\delta \ge -10^{\circ}$. 
    The center line represents perfect agreement while the outer lines 
    show an RMS combined error of 20\% (photometric distance error) and
    10 mas (trigonometric parallax error).}\label{photcom}
  \end{figure}
 
 \clearpage

  \begin{figure}
  \epsscale{1.00}
  \includegraphics[angle=-90,scale=0.5]{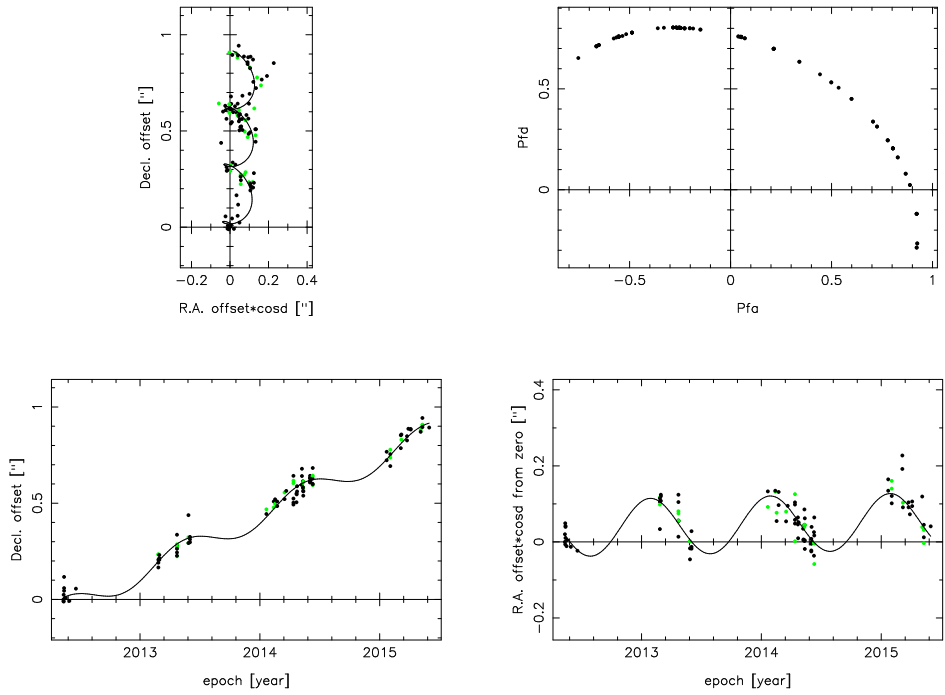}
  \caption{ Example of a good fit from our parallax pipeline.  This
    fit is for star G165-058 ($\pi$ = 80.1 $\pm$ 4.4 mas, 
    pmra = 6.4 $\pm$ 2.8 mas/yr, pmdc = 298.1 $\pm$ 2.7 mas/yr)
    showing in the top left
    the RA offset vs.~Dec offset [arcsecond], top right the
    parallactic ellipse, bottom left, Dec offset over time and
    bottom right RA offset over time.}\label{goodfit}
  \end{figure}

 \clearpage

  \begin{figure}
  \epsscale{1.00}
  \includegraphics[angle=-90,scale=0.5]{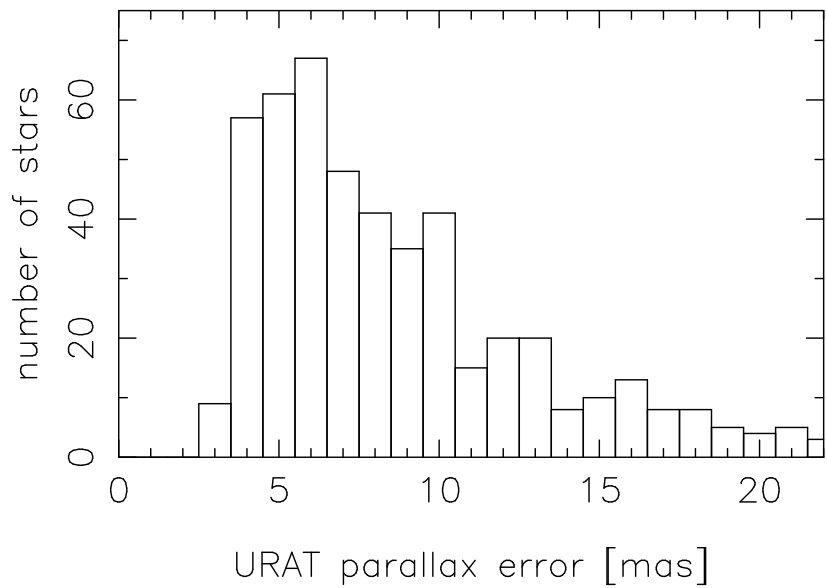}
  \caption{Histogram of the URAT parallax errors for the 545 stars
    having previous photometric distance estimates north of $\delta =
    -10^{\circ}$. }\label{pi-pre-perr}
  \end{figure}

 \clearpage

  \begin{figure}
  \epsscale{1.00}
  \includegraphics[angle=-90,scale=0.5]{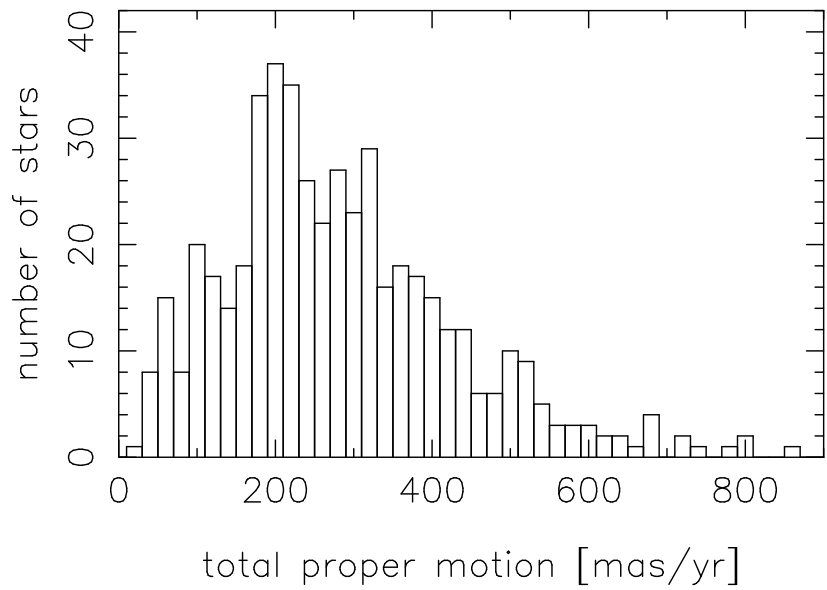}
  \caption{Histogram of the URAT total proper motion for the 545 stars
    having previous photometric distance estimates north of $\delta =
    -10^{\circ}$. }\label{pi-pre-tpm}
  \end{figure}
 
 \clearpage

  \begin{figure}
  \epsscale{1.00}
  \includegraphics[angle=-90,scale=0.5]{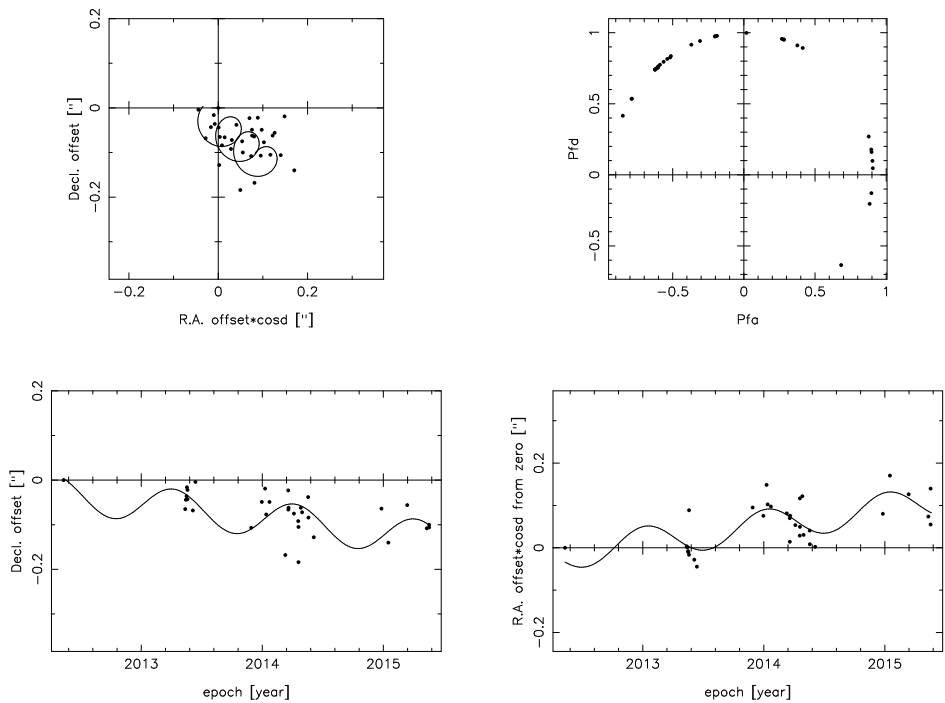}
  \caption{ Example of a poor fit from our parallax pipeline.  This fit
    is for UPM 1304+8611 ($\pi$ = 42.5 $\pm$ 7.3 mas, pmra = 40.1 $\pm$ 
    8.6 mas/yr, pmdc = -33.4 $\pm$ 8.4 mas/yr), which is marked as suspect
    in our table.  This plot shows all URAT epoch data for this star
    similarly to Figure~\ref{goodfit}.}\label{suspectfit}
  \end{figure}

 \clearpage

  \begin{figure}
  \epsscale{1.00}
  \includegraphics[angle=-90,scale=0.5]{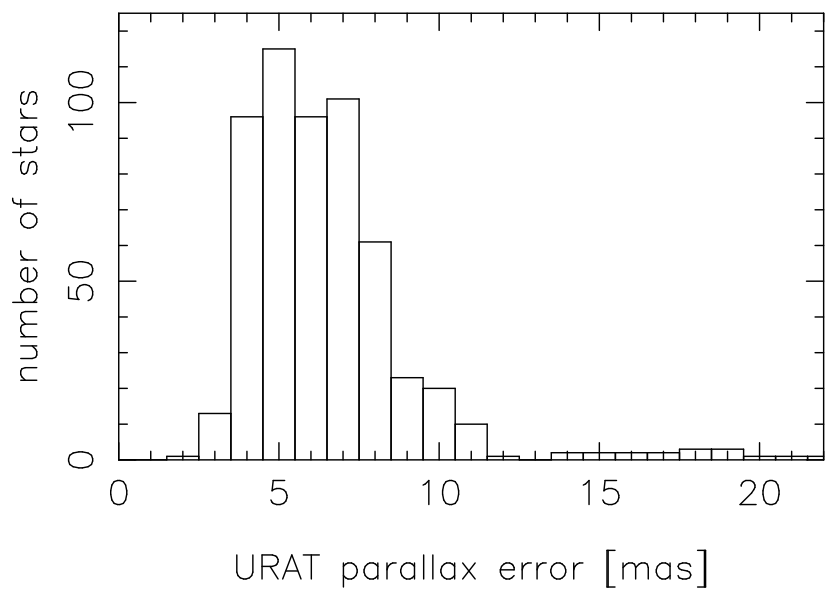}
  \caption{Histogram of the URAT parallax errors for the 558 new
    discoveries in our sample.}\label{pi-ndw-err}
  \end{figure}

 \clearpage

  \begin{figure}
  \epsscale{1.00}
  \includegraphics[angle=-90,scale=0.5]{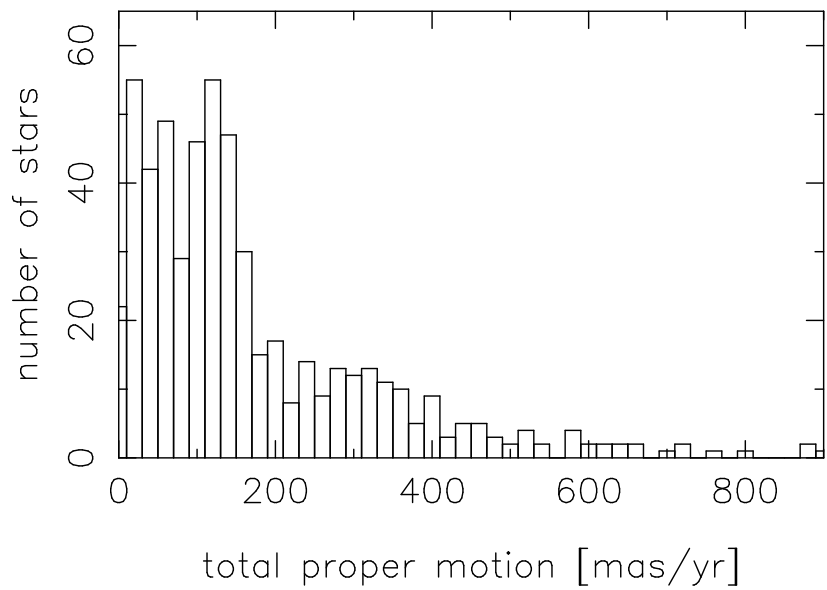}
  \caption{Histogram of the URAT total proper motion for the 558 new
    discoveries in our sample.} \label{pi-ndw-tpm}
  \end{figure}

 \clearpage

  \begin{figure}
  \epsscale{1.00}
  \includegraphics[angle=-90,scale=0.5]{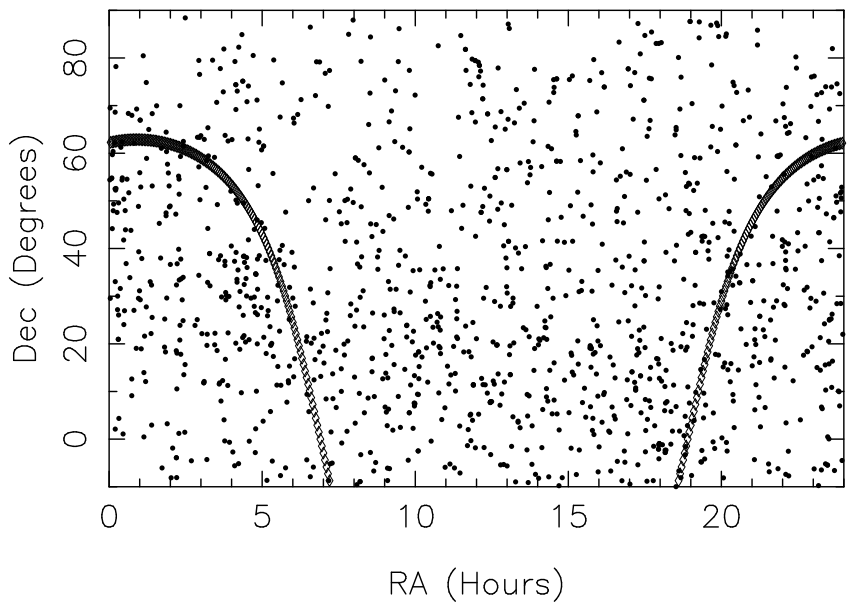}
  \caption{Distribution on the sky of all 1103 stars reported 
     in this paper. }\label{sky}
  \end{figure}
 
 \clearpage



\input{table_cuts.tex}

\clearpage
\input{table_err.tex}
\clearpage
\input{table_predistw2.tex}

\clearpage
\input{table_newdistw2.tex}
\clearpage


\end{document}

%% file: table_cuts.tex
\begin{deluxetable}{lrr}
\tabletypesize{}

\tablecaption{Adopted cuts for the URAT epoch data\label{table_cuts}}

\tablewidth{0pt}
\tablehead{
\colhead{CUT ITEM}            &
\colhead{LOW}                 &
\colhead{HIGH}                }

\startdata

FWHM               & \nodata &     7.0 \\    
Amplitude          &   500   &   30000 \\
sigma x [mas]      & \nodata &    90.0 \\
sigma y [mas]      & \nodata &    90.0 \\
numb. observ. used &    10   & \nodata \\
epoch span [year]  &   0.5   & \nodata \\

\enddata
\end{deluxetable}

%% file: table_err.tex
\begin{deluxetable}{lrr}
\tabletypesize{}

\tablecaption{Summary of URAT formal error investigation \label{table_err}}

\tablewidth{0pt}
\tablehead{
\colhead{ITEM}                   &
\colhead{case 1}                 &
\colhead{case 2}                 }

\startdata

low limit number of observations  &  20     & 30    \\
upper limit error URAT par. [mas] &  20.0   & 12.0  \\
upper limit error Hip. par. [mas] &  4.0    &  3.0  \\
lower limit epoch span [year]     &  1.5    & 2.0   \\
number of stars in sample         &  520    & 357   \\
upper limit par.difference [mas]  &  30.0   & 25.0  \\
number of stars remaining         &  511    & 347   \\
\hline
RMS of parallax difference [mas]  &  11.3   & 10.2  \\
RMS of formal errors [mas]        &  10.2   &  8.4  \\

\enddata
\end{deluxetable}

%% file: table_predistw2.tex
\begin{deluxetable}{lrrrcrrrrcrrrrrrrrrrrrl}
\vspace{-2.0in}
\rotate 
\tabletypesize{\tiny}
\setlength{\tabcolsep}{0.02in}

\tablecaption{New trigonometric parallaxes for stars having previous
  photometric distance estimates \label{table_predist}}

\tablewidth{0pt}
\tablehead{
\colhead{name}                 &
\colhead{R.A.}                 &
\colhead{decl.}                &
\colhead{Umag}                 &
\colhead{epdiff}               &
\colhead{obs.}                 &
\colhead{rej.}                 &
\colhead{$\Pi$(abs)}           &
\colhead{$\Pi$ err}            &
\colhead{$\Pi$(corr)}          &
\colhead{PMRA}                 &
\colhead{error}                &
\colhead{PMDC}                 &
\colhead{error}                &
\colhead{J}                    &
\colhead{H}                    &
\colhead{Ks}                   &
\colhead{B}                    &
\colhead{V}                    &
\colhead{g}                    &
\colhead{r}                    &
\colhead{i}                    &
\colhead{notes}               \\

\colhead{     }                &
\colhead{[deg]}                &
\colhead{[deg]}                &
\colhead{[mag]}                &
\colhead{[years]}              &
\colhead{[num]}                &
\colhead{[num]}                &
\colhead{[mas]}               &
\colhead{[mas]}               &
\colhead{[mas/y]}           &
\colhead{[mas/y]}           &
\colhead{[mas/y]}           &
\colhead{[mas/y]}           &
\colhead{[mas/y]}           &
\colhead{mag}                  &
\colhead{mag}                  &
\colhead{mag}                  &
\colhead{mag}                  &
\colhead{mag}                  &
\colhead{mag}                  &
\colhead{mag}                  &
\colhead{mag}                  &
\colhead{     }                }

\startdata

TYC 3980-1081-1           &   327.9090744 &    59.2944419 &  10.81 &  1.96 &  39 &   0 &  154.8 &   12.1 &    1.9 &  -21.0 &   13.6 &  149.5 &   13.7 &   6.529 &   5.860 &   5.651 &  11.958 &  10.501 &  11.185 &   9.980 &   8.600 &                    \\
LHS 6167                  &   138.9001086 &   -10.5972033 &  11.66 &  1.13 &  35 &   0 &  134.9 &   12.1 &    0.9 & -393.7 &   14.7 & -200.8 &   14.7 &   8.605 &   8.074 &   7.733 &  15.643 &  13.804 &  14.569 &  13.133 &  11.093 &                    \\
TYC 2187-512-1            &   320.5266342 &    22.9319081 &   9.39 &  1.31 &  20 &   0 &  109.3 &   19.1 &    1.0 &  131.4 &   28.1 &  120.0 &   20.0 &   7.400 &   6.800 &   6.595 &  12.021 &  10.555 &  11.336 &   9.980 &   9.103 &                    \\
LP688-003                 &   263.5356092 &    -8.8323381 &  13.82 &  2.10 &  18 &   0 &  108.0 &   21.8 &    3.9 & -171.5 &    7.7 & -293.1 &    7.7 &  10.905 &  10.400 &  10.055 & \nodata & \nodata & \nodata & \nodata & \nodata & \tablenotemark{a}  \\
TYC 3251-1875-1           &     3.9121606 &    47.5893683 &   9.59 &  2.12 &  58 &   0 &  104.9 &    9.6 &    1.2 &  -95.8 &   11.7 &   -5.5 &   11.8 &   7.336 &   6.709 &   6.467 &  12.605 &  11.120 &  11.804 &  10.511 &   9.234 &                    \\
2MASS J07171706-0501031   &   109.3229025 &    -5.0193367 &  11.47 &  2.24 &  51 &   0 &  104.6 &    7.9 &    1.3 &  424.7 &    5.8 & -396.4 &    5.7 &   8.873 &   8.349 &   8.045 &  14.716 &  13.224 &  13.928 &  12.608 &  11.067 &                    \\
G 184-31                  &   282.4765197 &    18.6735653 &  12.18 &  3.04 &  82 &   0 &  104.4 &    7.2 &    1.5 &  -97.8 &    4.0 & -277.7 &    3.9 &   9.380 &   8.815 &   8.509 &  14.881 &  13.619 &  14.209 &  13.109 &  11.702 &                    \\
\enddata

\tablenotetext{a} {large correction so cut off of 3.9 mas used to convert from relative to absolute parallax}
\tablenotetext{b} {no correction so mean of 1.3 mas used to convert from relative to absolute parallax}

\end{deluxetable}

%% file: table_newdistw2.tex
\begin{deluxetable}{lrrrcrrrrcrrrrrrrrrrrrl}
\rotate 
\tabletypesize{\tiny}
\setlength{\tabcolsep}{0.02in}

\tablecaption{New Nearby Star discoveries \label{table_newdist}}

\tablewidth{0pt}
\tablehead{
\colhead{name}                 &
\colhead{R.A.}                 &
\colhead{decl.}                &
\colhead{Umag}                 &
\colhead{epdiff}               &
\colhead{obs.}                 &
\colhead{rej.}                 &
\colhead{$\Pi$(abs)}           &
\colhead{$\Pi$ err}            &
\colhead{$\Pi$(corr)}          &
\colhead{PMRA}                 &
\colhead{error}                &
\colhead{PMDC}                 &
\colhead{error}                &
\colhead{J}                    &
\colhead{H}                    &
\colhead{Ks}                   &
\colhead{B}                    &
\colhead{V}                    &
\colhead{g}                    &
\colhead{r}                    &
\colhead{i}                    &
\colhead{notes}                \\

\colhead{     }                &
\colhead{[deg]}                &
\colhead{[deg]}                &
\colhead{[mag]}                &
\colhead{[years]}              &
\colhead{[num]}                &
\colhead{[num]}                &
\colhead{[mas]}                &
\colhead{[mas]}                &
\colhead{[mas]}                &
\colhead{[mas/y]}        &
\colhead{[mas/y]}        &
\colhead{[mas/y]}        &
\colhead{[mas/y]}        &
\colhead{mag}               &
\colhead{mag}               &
\colhead{mag}               &
\colhead{mag}               &
\colhead{mag}               &
\colhead{mag}               &
\colhead{mag}               &
\colhead{mag}               &
\colhead{     }                }

\startdata

UPM0429+3806              &    67.3570786 &   +38.1026178 &  15.36 &  2.35 &  39 &   3 &    227.2 &     19.0 &      2.8 &    114.7 &     14.0 &      3.8 &     13.0 &  13.636 &  13.204 &  13.152 &  16.568 &  15.455 &  16.044 &  15.216 &  14.717 &                  \tablenotemark{a}\tablenotemark{b}  \\
UPM2340+4625              &   355.0318225 &   +46.4188983 &  14.93 &  2.21 &  47 &   1 &    206.5 &     17.3 &      0.9 &     48.5 &     12.3 &     42.0 &     12.0 &  13.658 &  13.232 &  13.137 & \nodata & \nodata & \nodata & \nodata & \nodata &                  \tablenotemark{a}\tablenotemark{b}  \\
UPM2340+4631              &   355.0836800 &   +46.5196811 &  15.61 &  2.21 &  33 &   0 &    177.9 &     24.1 &      0.9 &     18.6 &     16.2 &     31.0 &     16.2 &  14.110 &  13.611 &  13.492 & \nodata & \nodata & \nodata & \nodata & \nodata &                  \tablenotemark{a}\tablenotemark{b}  \\
UPM0416+3100              &    64.1374097 &   +31.0006092 &  14.50 &  2.35 &  52 &   0 &    176.3 &     17.9 &      1.3 &     43.2 &     10.9 &     -8.7 &      9.9 &  13.086 &  12.645 &  12.605 &  15.722 &  14.728 &  15.189 &  14.382 &  14.014 &                  \tablenotemark{a}\tablenotemark{b}  \\
UPM0012+5215              &     3.1964658 &   +52.2626775 &  15.62 &  2.21 &  34 &   1 &    154.3 &     17.3 &      1.1 &     42.6 &     14.5 &     46.2 &     14.0 &  14.374 &  14.049 &  14.075 & \nodata & \nodata & \nodata & \nodata & \nodata &                                   \tablenotemark{b}  \\
TYC 3576-1996-1           &   303.7610778 &   +47.8085700 &  10.91 &  2.01 &  45 &   8 &    142.3 &     18.7 &      1.4 &    -77.0 &     18.5 &     11.4 &     15.3 &   6.274 &   5.373 &   5.114 &  11.958 &   9.902 & \nodata & \nodata & \nodata &                  \tablenotemark{a}\tablenotemark{b}  \\
\enddata

\tablenotetext{a} {visual inspection of the fit shows scatter indicating the solution might be erroneous}
\tablenotetext{b} {visual inspection of the sky image shows an elongated or blended image indicating the solution might be erroneous}
\tablenotetext{c} {no correction so mean of 1.3 mas used to convert from relative to absolute parallax}
\tablenotetext{d} {large correction so cut off of 3.9 mas used to convert from relative to absolute parallax}

\end{deluxetable}